# P2DFlow: A Protein Ensemble Generative Model with SE(3) Flow Matching

*Yaowei Jin[1], Qi Huang[5], Ziyang Song[6], Mingyue Zheng[2,3,4], Dan Teng[2, *], Qian Shi[1, *]*

1 Lingang Laboratory, Shanghai 200031, China.

2 Drug Discovery and Design Center, State Key Laboratory of Drug Research, Shanghai Institute of Materia Medica, Chinese Academy of Sciences, 555 Zuchongzhi Road, Shanghai 201203, China.

3 University of Chinese Academy of Sciences, No.19A Yuquan Road, Beijing 100049, China.

4 School of Chinese Materia Medica, Nanjing University of Chinese Medicine, Nanjing 210023, China.

5 Institute for Electric Light Sources, School of Information Science and Technology, Fudan University, Shanghai 200438, P. R. China.

6 Shanghai Key Lab of Chemical Assessment and Sustainability, School of Chemical Science and Engineering, Tongji University, Shanghai 200092, P. R. China.

*To whom correspondence should be addressed.

*(Qian Shi) E-mail: shiqian@lglab.ac.cn

*(Dan Teng) E-mail: tengdan@simm.ac.cn

# Abstract

Biological processes, functions, and properties are intricately linked to the ensemble of protein conformations, rather than being solely determined by a single stable conformation. In this study, we have developed P2DFlow, a generative model based on SE(3) flow matching, to predict the structural ensembles of proteins. We specifically designed a valuable prior for the flow process and enhanced the model's ability to distinguish each intermediate state by incorporating an additional dimension to describe the ensemble data, which can reflect the physical laws governing the distribution of ensembles, so that the prior knowledge can effectively guide the generation process. When trained and evaluated on the MD datasets of ATLAS, P2DFlow outperforms other baseline models on extensive experiments, successfully capturing the observable dynamic fluctuations as evidenced in crystal structure and MD simulations. As a potential proxy agent for protein molecular simulation, the high-quality ensembles generated by P2DFlow could significantly aid in understanding protein functions across various scenarios. Code is available at https://github.com/BLEACH366/P2DFlow.

# 1. Introduction

Proteins exhibit a dynamic nature that leads to the generation of diverse conformations. Crucial biological functions are executed by relying on the distinct states, collective motions, and disordered fluctuations within the protein ensemble[1, 2]. Thus, understanding the distribution of these ensembles is essential for elucidating the mechanism by which proteins function in different environments. While experimental measurements, such as crystallographic B-factors and NMR spectroscopy, can provide some insight into conformational changes, they are limited in spatial and temporal scale[3, 4]. Although methods like AlphaFold[5], ESMFold[6], and other deep learning approaches have demonstrated excellent performance in predicting the crystal structure of proteins, it is still challenging to offer diverse predictions for protein conformational ensembles [7-10].

Several computational methods are available for conformational sampling. Traditional approaches include Monte Carlo (MC) and molecular dynamics (MD). By providing an initial structure of a molecular, these methods explore its conformational space based on the forces acting upon it, which can be calculated using molecular mechanics or quantum mechanics[11]. However, these methods face several challenges: the computational efficiency declines rapidly as the number of atoms and degrees of freedom increase, and both MC and MD are reliant on force fields and energy, making it difficult to overcome high energy barriers. Consequently, these methods often become trapped in local minima. To address these issues, enhanced sampling methods, such as umbrella sampling [12] and meta dynamics [13], are employed for broader exploration,.

An alternative approach involves the use of generative models leveraging machine learning and deep learning techniques. For instance, Boltzmann generator [14] represents one of the earliest attempt to use normalizing flow to sample system-specific conformational distributions from random noise. However, it requires pre-acquired simulation data for specific protein systems. Multiple sequence alignment (MSA) subsampling combined with AlphaFold [15-17] can predict different structures for each MSA subset. While effective in certain contexts, it faces two limitations. First, the results depend on the partition of the MSA, making it difficult to predict protein ensembles lacking homologous proteins. Furthermore, it does not provide the approach for training the model on protein ensemble data, which is critically important. DiG[18] integrates diffusion models with a simulated annealing process to generate diverse molecular conformations while providing state density estimation and property-guided generation. It is applicable to macromolecular proteins, organic systems, and inorganic catalytic systems. BioEmu[19] adopts a model architecture similar to DiG, leveraging large-scale protein structure data for pretraining and fine-tuning with protein stability data derived from MD simulations and experimental measurements. This approach enables BioEmu to qualitatively sample a wide range of functionally relevant conformational changes and predict protein stability. DeepConformer[20] leverages experimental structures from the Protein Data Bank and co-evolutionary data to learn protein energy landscapes and conformational dynamics without relying on MD simulation data. By modifying multiple sequence alignments (MSA) and incorporating amino acid masking techniques, it enhances the model's ability to capture a comprehensive protein energy landscape. STR2STR [4] employs a diffusion model to learn the score matching of distributions from random noise to protein ensemble distributions, combining stochastic perturbation and score to guide the direction of conformational changes. It does not rely on simulation data during either training or inference and is capable of performing zero-shot conformation sampling. However, the large variability in its predicted results makes it difficult to focus on a particular low-energy conformation. This may be due to inconsistencies in its prior distribution during training and sampling processes. AlphaFlow [21] integrates AlphaFold and ESMFold with flow matching, fine-tune these models to convert regression models to generative models. It can generate accurate and diverse PDB structures, which can be utilized to compute additional dynamic properties. Nonetheless, it encounters the challenges in generating non-existent intermediate states between two or more minima, similar to the issues faced by AlphaFold in certain proteins.

To address these limitations, we propose a new framework that samples protein ensembles via a SE(3) equivariant flow matching model named P2DFlow (Fig. 1). P2DFlow is trained on MD simulation data, which contains different macro states of proteins (as seen in Fig. 4), allowing it to learn the conformational changes within protein ensembles. To enhance the model's generalization ability while reducing the difficulty of training, we use a strong prior, ESMFold predictions with perturbations of coordinates (detailed in Sec. 2.1), to obtain an estimated structure. Compared to Gaussian prior and harmonic prior, the stronger prior that we use guides the model to generate accurate structures by introducing more precise biases in bond lengths and bond angles. This approach differs significantly from AlphaFlow, which uses harmonic prior and relies on AlphaFold or ESMFold as the main component to control structure. In contrast, P2DFlow employs ESMFold predictions with perturbations as a stronger prior and utilizes a new SE(3) equivariant block to adjust the structure. To distinguish between various conformations of the same protein, we introduce a new dimension called 'approximate energy', which maps MD results onto a low-dimensional plane and uses the density to represent probability (detailed in Sec. 2.2). It guides the model in generating conformations with distinct energies, thereby avoiding the generation of non-existent intermediate states.

We evaluate the performance of P2DFlow against AlphaFlow and STR2STR, two representative and advanced models for protein ensemble generation. Based on extensive experiments on the test dataset of ATLAS[22], it is evident that P2DFlow outperforms other baseline models on the metrics which can reflect the fidelity and dynamic properties of the generated ensembles compared to the ground truth. Visualization results indicate that P2DFlow effectively captures important changes in residue contact and more accurately recovers protein ensembles distributions. Moreover, the ablation experiment shows that the 'approximate energy' significantly aids in constructing the conformation distribution of proteins. Thus, P2DFlow can be used to predict protein ensembles without the need for expensive MD simulations.

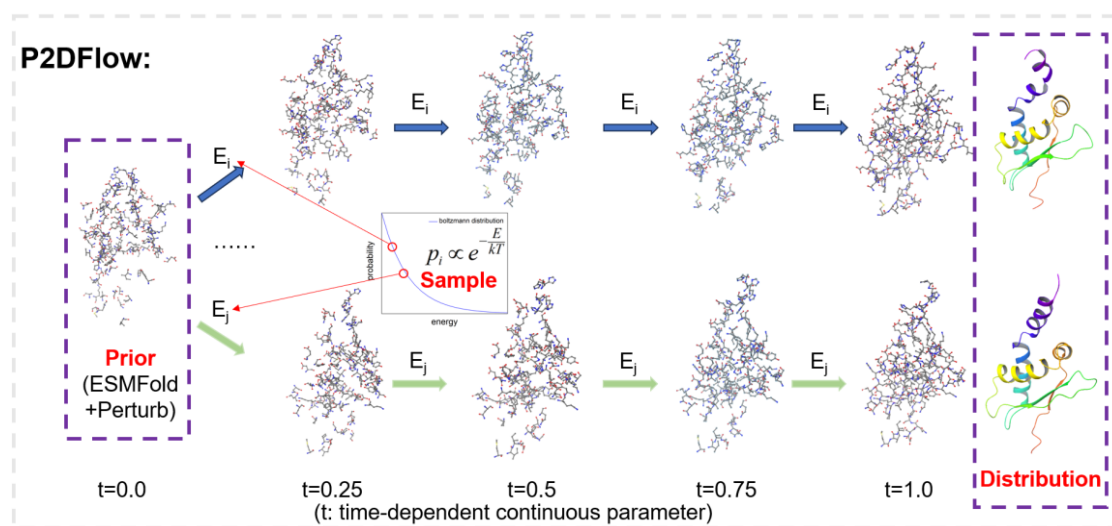

**Figure 1. Conceptual overview of P2DFlow.** Protein structures are firstly sampled from a prior based on perturbed ESMFold predictions. Subsequently, an 'approximate energy' sampling serves as a condition to guide the generation process to produce ensemble distribution.

## 2. Method

### 2.1 SE(3) Flow Matching

A flow-matching generative framework[23] aims to learn continuous normalizing flows (CNFs) [24], generating target distributions by predicting a vector field and integrating the ordinary differential equation (ODE).

For a flow $\varphi(x, t)$, there is a related vector field $v(x, t)$, defined by the following ODE:

$$\frac{d\varphi(x,t)}{dt} = v(x,t), \qquad \varphi(x,0) = x_0 \tag{1}$$

where $t \in [0, 1]$ is a continuous parameter, $x_0$ is a sample of prior distribution $p_0$. We can use the flow to transform the simple prior $p_0$ towards the data distribution $p_1$ by integrating the vector field $v(x, t)$:

$$x_1 - x_0 = \int_0^1 v(x,t)\, dt \tag{2}$$

For the flow, since it only fixes the distribution at $t = 1$ and the prior at $t = 0$, the interpolation process between them can be manually defined. In our method, we choose ESMFold prediction with perturbations as the prior. It means that, for a given protein sequence requiring ensemble generation, we first use ESMFold to predict the static structure. Then, Gaussian noise with specific variance is added to the coordinates of each residue to generate the prior distribution. Previous studies has shown that the prior distribution closer to data distribution which needs to be learned can lead to superior performance[25], so we identify this as an especially valuable inductive bias since it will offer a suitable bond length and torsion angle as the initial value of the iteration, compared with Gaussian prior and harmonic prior [26, 27].

For the interpolation process, following previous work [28-30], the backbone atom positions of each residue are parameterized by an orientation preserving rigid transformation $T \in SE(3)$, that maps from fixed coordinates of $N^*$, $C_\alpha^*$, $C^*$, $O^*$ centered at $C_\alpha^* = (0,0,0)$. Each frame $T = (r, x)$ consists of a rotation $r \in SO(3)$ and a translation vector $x \in R^3$. As for the side chain of the residue, torsion angles $\theta$ are used to define the twists. Since our focus lies on the backbone of protein, we only interpolate for translation and rotation accompanied by the use of optimal transport (OT) path [31]:

$$Translations\ (R^3): x_t = (1-t)x_0 + tx_1 \quad (3)$$

$$Rotations\ (SO(3)): r_t = \exp_{r_0}(t \log_{r_0}(r_1)) \quad (4)$$

We calculate $\exp_{r_0}$ and $\log_{r_0}$ using Rodrigues' formula [32]. The vector field for translation and rotation are then computed as follow:

$$\dot{x}_t = \frac{x_1 - x_t}{1 - t}, \qquad \dot{r} = \frac{\log_{r_t}(r_1)}{1 - t} \quad (5)$$

These vector fields are learned with SE(3) equivariant neural network (mentioned in Section 2.2). For the torsion angle within residues, we predict it using the node embedding of the corresponding residues with a Multi-Layer Perceptron (MLP) due to its SE(3) invariant[33], which is similar to the approach utilized in AlphaFold.

With these considerations and calculations mentioned above, the training loss of P2DFlow is defined as:

$$L = E_{t,p_1,p_0}\{\frac{1}{(1-t)^2}(\|\hat{x}_1 - x_1\|^2 + \|\hat{r}_1 - r_1\|^2)$$

$$+ \alpha(\|\hat{C}_1(\hat{x}_1, \hat{r}_1, \hat{\theta}) - C_1\|^2 + \|\hat{D}_1(\hat{x}_1, \hat{r}_1, \hat{\theta}) - D_1\|^2)\} \quad (6)$$

Where $(\hat{x}_1, \hat{r}_1, \hat{\theta})$ refer to predicted translation, rotation, and torsion at $t = 1$. $(\hat{C}_1, \hat{D}_1)$ refer to the atom coordinates and distance matrix of all atoms, which is recovered using $(\hat{x}_1, \hat{r}_1, \hat{\theta})$ and standard frames proposed in AlphaFold at $t = 1$. $\alpha$ is a hyperparameter to adjust the auxiliary loss with the vector field loss. Furthermore, we subtract the Center of Mass (CoM) from the prior sample and each protein data to achieve a zero CoM. This step is crucial for ensuring that the distribution of the sampled results remains SE(3) equivariant.

## 2.2 Network Architecture

To learn the aforementioned vector field, we developed P2DFlow, a flow-matching model that is SE(3) equivariant. The model's equivariance is achieved by the stacking of Invariant Point Attention (IPA) module [5] and E(n) equivariant Graph Neural Network (EGNN) module [34]. These components are utilized to encode the spatial features of the protein. In the molecular graph representation, residues are considered as nodes, and edges are defined by two criteria: the cutoff coordinate distance between adjacent residues and the cutoff sequence distance. We utilize the union of these criteria to establish the edges.

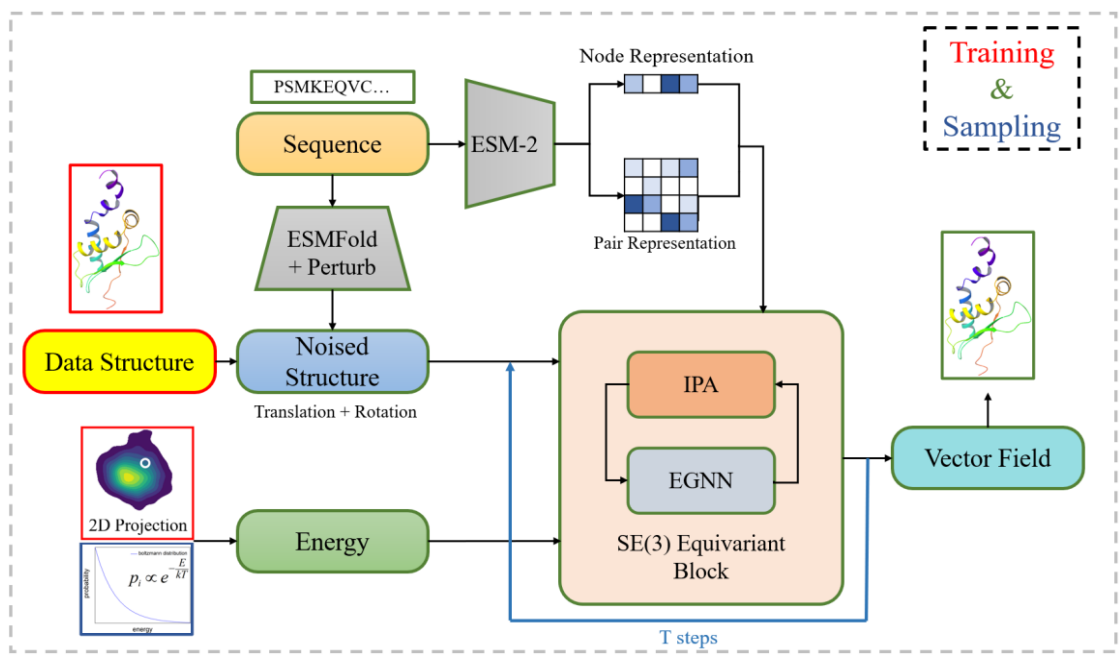

**Figure 2. Workflow of P2DFlow.** Protein sequence and 'approximate energy' are inputs of P2DFlow. We utilize ESM-2 to get embeddings from sequence, and use ESMFold + Perturb to get the noised structure as the prior distribution, then apply a SE(3) equivariant block composed with IPA and EGNN to predict the vector filed for the flow process. The differences between training and sampling are marked with different border colors.

The inputs of the model include sequence and 'approximate energy' (Figure 2). To improve the generalization ability of the model, we utilize the ESM-2 protein language model to generate the initial Node Representation (Node Repr.) and Pair Representation (Pair Repr.) for residues. After successive aggregations and updates of the Node Repr. and Pair Repr. using the SE(3) Equivariant Block, we obtain a compressed, full-graph representation of the protein structure, which is sampled from prior of ESMFold prediction with perturbation. Subsequently, we employ the representation to predict the vector fields associated with translation and rotation.

To distinguish structures with different energies in the conformation set, we introduce the concept of 'approximate energy'. We project the molecular dynamics (MD) simulation ensembles onto a two-dimensional plane defined by the radius of gyration (RG) and the root-mean-square deviation (RMSD) relative to the crystal structure. We then compute the Gaussian kernel density of this 2D map and apply the Boltzmann Equation to convert the density values into 'approximate energy' values after normalization.

The sampling process is illustrated in Figure 2. Given a protein sequence as input, the process begins with the prediction of a stable structure using ESMFold. Subsequently, we add Gaussian noise with the same variance used during training to perturb the structure, so that we can get the initial perturbed structure from the specialized prior distribution. We then sample the 'approximate energy' from Boltzmann Distribution, and keep it unchanged throughout the entire autoregressive sampling process to guide the generation. Following the prediction of vector fields, a Euler integration step is applied to solve the ODE. Ultimately, we reconstruct the coordinates of each atom based on the translation, rotation, and torsion angles, utilizing the frame representation provided by AlphaFold, to derive the protein conformation structure. After conducting a sufficient number of samples, we obtain the predicted protein ensembles.

Other operational details are in line with those of FrameFlow [30]. We modify the loss function by substituting $1/(1-t)^2$ with $1/(1-min\{t, 0.9\})^2$ to prevent the loss from exploding. Furthermore, we conduct a pre-alignment[35, 36] to align $x_0$ and $x_1$ using the Kabsch algorithm. Specifically, we solve $r^* = argmin_{r \in SO(3)} \|rx_0 - x_1\|^2$ and use the aligned position $r^* x_0$ during training.

## 3. Experiments

### 3.1 Setup

**Training** To learn the dynamic distribution of protein ensembles, we utilize ATLAS dataset[22], which contains ~1300 MD simulation results for proteins. To select representative structures from the conformations, we compute the 'approximate energy' (mentioned in section 2.2) for each of them, then uniformly select 100 representative structures based on the values of 'approximate energy' to increase the proportion of representative structures that are distinct from the low-energy crystal structure. For evaluation, we randomly choose ~100 ensembles from ATLAS dataset excluding the training set. Our model is trained for 8 days using 4 NVIDIA A100-80G GPUs.

**Metrics** To assess the performance of P2DFlow, we use evaluation metrics which can be categorized into: (a) **Validity**. Assesses whether the sampled conformations obey basic physical constraints. This is defined by the ratio of conformations that pass the sanity check, which includes criteria such as the absence of steric clashes (Steric), broken bonds (Bond) and the consistency of dihedral angle distributions in the context of the Ramachandran plot (RP). (b) **Fidelity**. Reflects the distributional gap between sampled ensemble and reference MD simulation, which contains Jensen-Shannon divergence of pairwise distance[4] (PWD $\mathcal{J}$) and radius of gyration (RG $\mathcal{J}$), root mean Wasserstein distance (RMWD $\mathcal{W}_2$) and root mean square fluctuation (RMSF). (c) **Dynamics**. Shows the observable structural change which is often associated with thermal fluctuations, such as the Weak Contacts $\mathcal{J}a$ and Transient Contacts $\mathcal{J}a$ [21]. Weak contacts are defined as Cα pairs that are in contact in the crystal structure but dissociate in more than 10% of the ensemble structures, while transient contacts are those not in contact in the crystal structure but associate in more than 10% of the ensemble structures.

**Baseline** We compare our results with Alphaflow and STR2STR. Alphaflow is a flow matching model based on fine-tuning of AlphaFold and ESMFold, and STR2STR is a diffusion model which changes inference stage by adding AlphaFold prediction as the initial structure. We use their pretrained weights offered at GitHub to sample ensembles on the test set of ~100 proteins from ATLAS dataset.

## 3.2 Results

## 3.2.1 Overview of the results on the test set

Table 1 shows the evaluation results for P2DFlow, AlphaFlow and STR2STR. For each method, we compare the predicted ensemble with the ground truth MD ensembles and evaluate them on various metrics. The results show that P2DFlow outperforms AlphaFlow and STR2STR across all three types of metrics. This indicates that P2DFlow more realistically reflect the physical space distribution of structures, resulting in generated ensembles that closely resemble the ground truth derived from MD simulation. As for the dynamics of protein residue contacts, P2DFlow can predict the residue pairs with a higher likelihood of dissociation, aiding the identification of potential protein pockets.

**Table 1. Evaluation on MD ensembles.** $\mathcal{J}$: Jensen-Shannon divergence; $\mathcal{W}_2$: 2-Wasserstein distance; $\mathcal{J}a$: Jaccard similarity. Among these metrics, Steric, Bond, RP, Weak Contacts $\mathcal{J}a$ and Transient Contacts $\mathcal{J}a$ are the higher the better (↑); while PWD $\mathcal{J}$, RG $\mathcal{J}$, RMWD $\mathcal{W}_2$, RMSF are the lower the better (↓). The best result from generative models is bolded.

|  |  | STR2STR | AlphaFlow | P2DFlow |
|---|---|---|---|---|
|  | Steric ($\uparrow$) | 0.667 | 0.725 | **0.941** |
| Validity | Bond ($\uparrow$) | 0.964 | 0.939 | **1.000** |
|  | RP ($\uparrow$) | 0.863 | 0.865 | **0.937** |
|  | PWD $\mathcal{J}$ ($\downarrow$) | 0.446 | 0.452 | **0.331** |
|  | RG $\mathcal{J}$ ($\downarrow$) | 0.461 | 0.519 | **0.429** |
| Fidelity | RMWD $\mathcal{W}_2$ ($\downarrow$) | 78.246 | 33.628 | **15.168** |
|  | RMSF ($\downarrow$) | 3.997 | 0.576 | **0.268** |
| Dynamics | Weak Contacts $\mathcal{J}a$ ($\uparrow$) | 0.329 | 0.458 | **0.710** |
|  | Transient Contacts $\mathcal{J}a$ ($\uparrow$) | 0.148 | 0.288 | **0.422** |

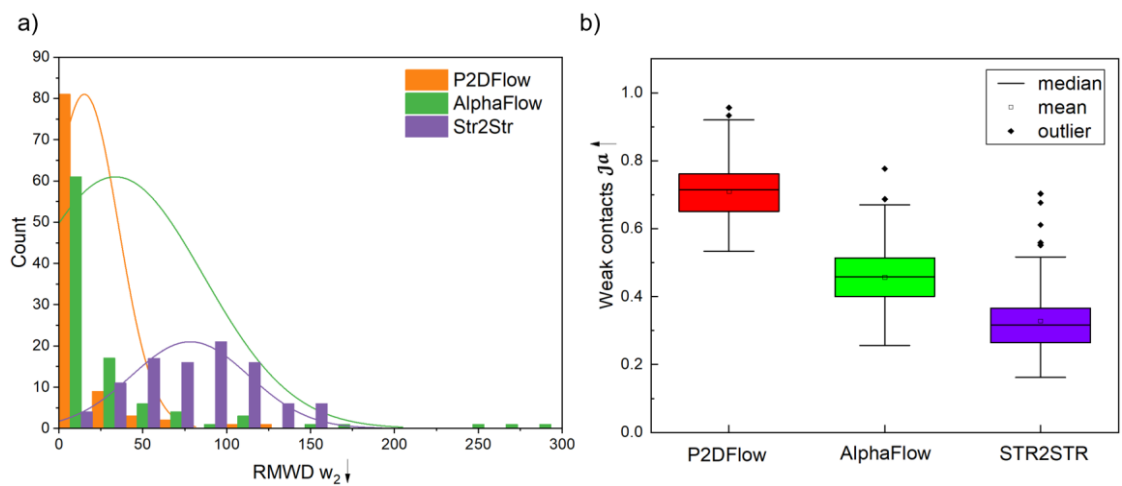

**Figure 3. Distribution of evaluation results for selected metrics. a)** RMWD $\mathcal{W}_2$. **b)** Weak Contacts $\mathcal{J}a$. The test set comprises approximately 100 types of protein ensembles, and bins were established for the metrics to generate the histograms. Subsequently, a Gaussian distribution was fitted to the histograms.

Figure 3 illustrates the distribution of RMWD $\mathcal{W}_2$ and Weak Contacts $\mathcal{J}a$ results on the ATLAS test set. We can see that there are a few bad outliers predicted by AlphaFlow and STR2STR, whereas P2DFlow does not exhibit such issues. Additionally, the Gaussian distribution of results generated by P2DFlow exhibits narrower peaks compared to other methods. This indicates more stable predictions with a lower likelihood of extreme outcomes across different protein ensembles. The box plot distribution further corroborates this observation.

### 3.2.2 Protein domain motion and conformational space coverage

Conformational transitions in proteins are critical for understanding their functional mechanisms, yet they remain challenging to capture accurately in molecular dynamics simulations and protein structure predictions. To enhance the robustness and biological relevance of our study, we have included some test cases that exemplify well-documented conformational transitions. In human physiological processes, the structural transitions of D-ribose binding protein (PDB IDs 2dri and 1urp) and adenylate kinase (PDB IDs 1ake and 4ake) play a critical role in regulating molecular transport and metabolic pathways. For the D-ribose binding protein, 1urp corresponds to the open state without ligand binding, whereas 2dri represents the closed state upon binding to D-ribose. This conformational transition plays a crucial role in the ABC (ATP-binding cassette) transport system, ensuring the efficient uptake and metabolic utilization of D-ribose. For adenylate kinase, 4ake corresponds to the open state without substrate binding, whereas 1ake represents the closed state upon binding to AMP. These two states directly affect cellular energy homeostasis and signal regulation.

To evaluate the model's ability to predict protein domain motion, we take proteins D-ribose binding protein and adenylate kinase as examples. According to the definition proposed in BioEmu [19] by Microsoft, we considered structures with RMSD < 3 Å as successfully samples. As shown in Figure 4a, P2DFlow achieved a 100% success rate for the D-ribose binding protein and successfully sampled both the 2dri and 1urp conformational states. However, a majority of the sampled results are closer to 2dri, which is attributed to the bias in the prior structure generated by ESMFold. A similar issue is also observed in the results of BioEmu. However, this problem can be mitigated by refining the sampling strategy for P2DFlow. Specifically, we can utilize an autoregressive approach, where the sampled result with the largest RMSD difference from the initial prior structure is used as the new prior for the next round of structure generation. As illustrated in Figure 4b, for adenylate kinase, by using the structure with the largest RMSD from 1ake in the 1st sampling round (marked as a white star) as the initial structure for the 2nd sampling round, we obtained two distinct cluster centers—one corresponding to 1ake and the other to 4ake. This approach effectively expands the model's coverage of the protein conformational space.

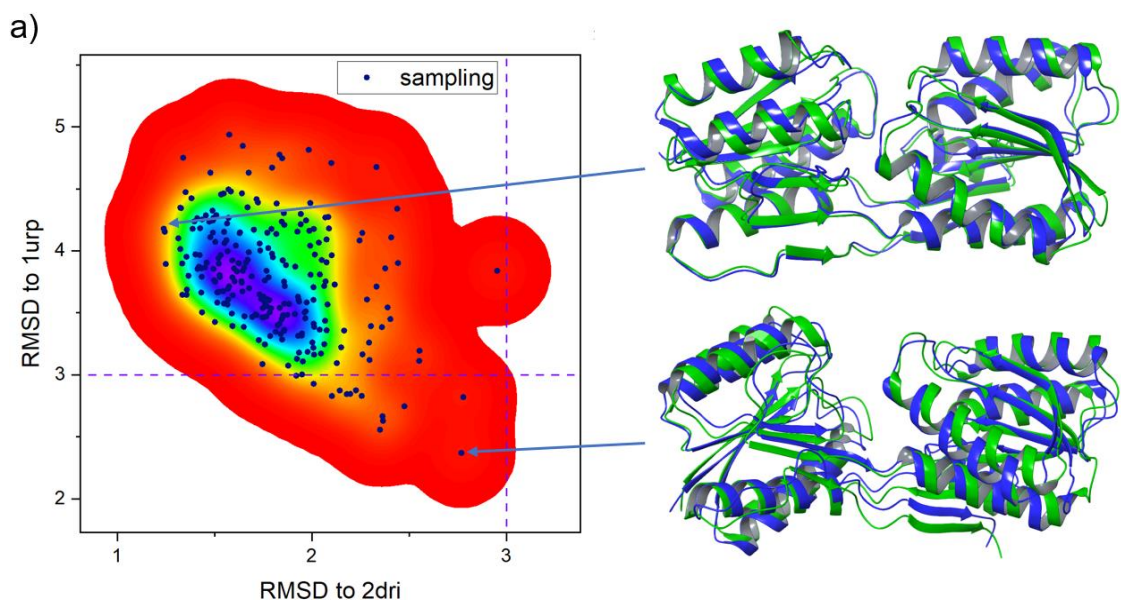

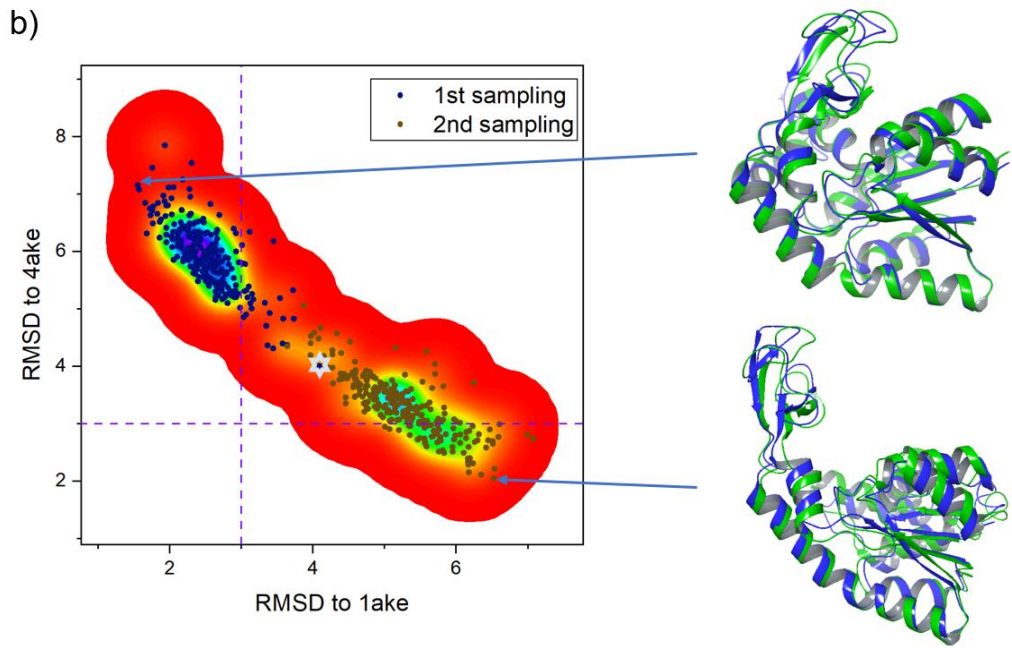

**Figure 4. Projection of protein ensembles subsampled from P2DFlow, plotted according to their backbone RMSDs relative to (a) D-ribose binding protein 2dri and 1urp (b) adenylate kinase 1ake and 4ake.** The experimental structures (blue) and their most similar (of the lowest RMSD) predicted structures (green) are shown on the right. The new initial structure for the 2nd sampling round in (b) is marked as a white star.

As for the Principal Component Analysis (PCA) of predictions in Figure 5, we can see that the predicted structures of P2DFlow can cover all the three cluster centers of the actual MD simulation, while AlphaFlow tends to predict structures only for two cluster centers, indicating challenges in generalizing across diverse conformations. STR2STR shows scattered predictions throughout the PCA plane, with some results falling outside the expected range, sacrificing accuracy for diversity. Compared to the results reported in the DiG [18] paper from Microsoft, the sampling results there tend to contain a higher proportion of high-energy structures that deviate significantly from the cluster center, based on the reduced 2D projection map. This issue necessitates a more complex post-processing procedure for selection and optimization.

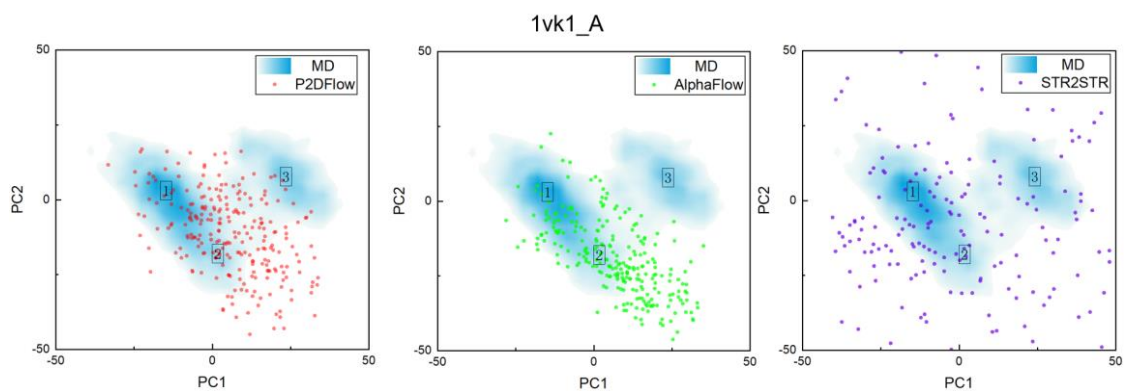

**Figure 5. PCA of MD ensembles and model predictions by P2DFlow, AlphaFlow and STR2STR for PDB ID 1vk1_A.** The centers of the clusters are marked with numerical labels.

### 3.2.3 Protein dynamic changes

For dynamic property evaluation, P2DFlow accurately predicts weak contacts of residues revealed by the crystal structure and MD simulations for 3rmq_A protein in the loop region (Figure 6a), suggesting the presence of potential pockets. Similarly, for the transient contacts of residues in 6jpt_A at the end of two beta-sheets (Figure 6b), P2DFlow demonstrates its ability to find instantaneous association interaction. Identifying these weak and transient contacts of residues from the generated ensembles aids in analyzing thermal fluctuations around the low-energy crystal structure.

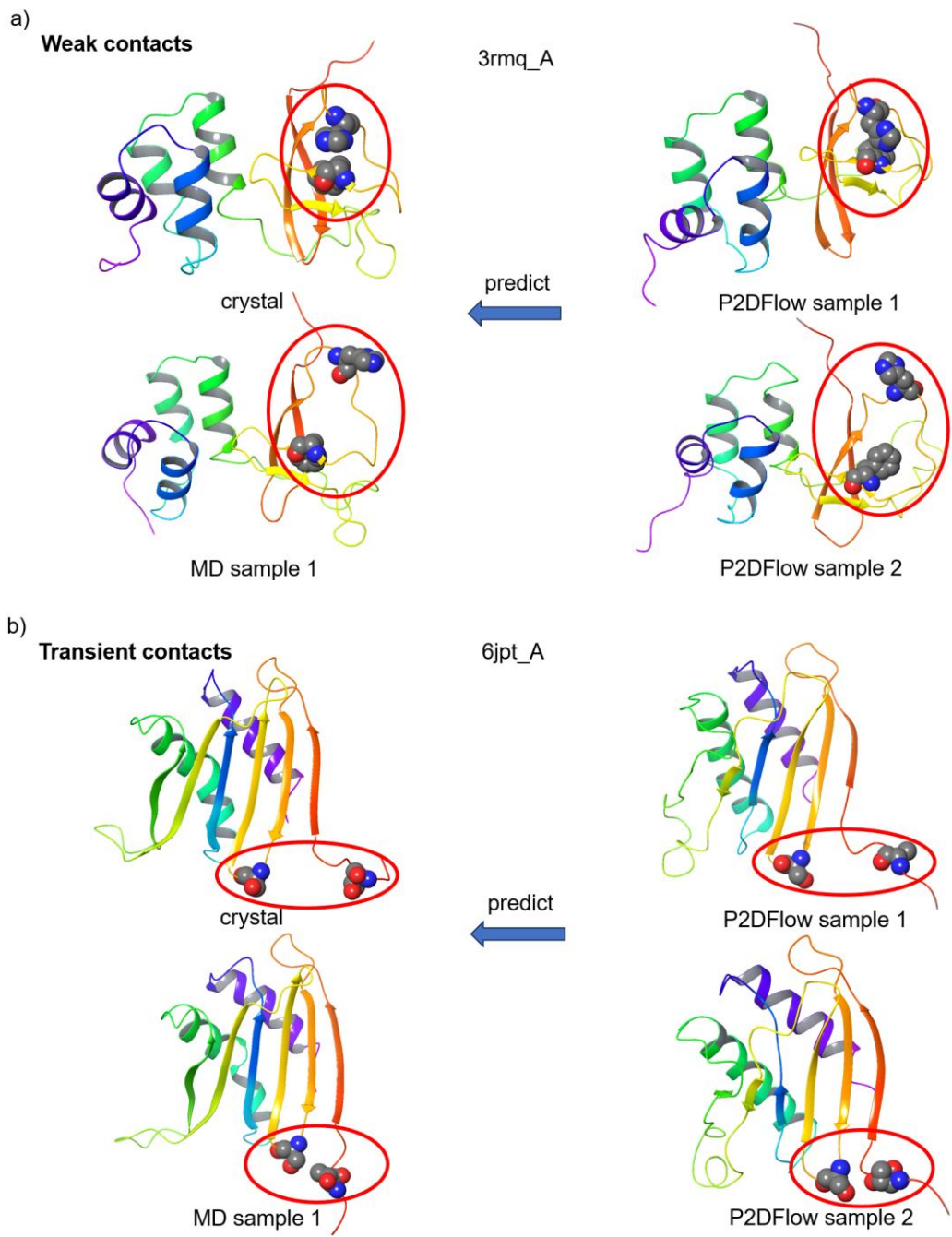

**Figure 6. Protein dynamic property evaluation. a**) dissociation of a weak residue contact in PDB ID 3rmq_A. **b**) association of a transient residue contact in PDB ID 6jpt_A.

By comparing the RMSF, the flexibility and fluctuation of individual residues in the protein structure can be evaluated. Here, we compare the predicted structure ensemble generated by P2DFlow with MD ensemble for 1g6g_A. As shown in the left panel of

Figure 7, P2DFlow accurately captures the variation in residue fluctuations, with nearly all peaks in the MD results being correspondingly reflected. The right panel of Figure 7 illustrates the structural alignment between the P2DFlow-predicted ensemble and the corresponding experimental structure. The red circle highlights the N-terminal β-sheet initiation, while the yellow circle marks the C-terminal α-helix. These circled regions exhibit greater variability in their extension directions and spatial positioning compared to other regions, which aligns well with RMSF profile. Compared to DeepConformer [20], whose RMSF plots in the paper exhibit a tendency to merge certain adjacent peaks into a single peak, P2DFlow provides more precise predictions.

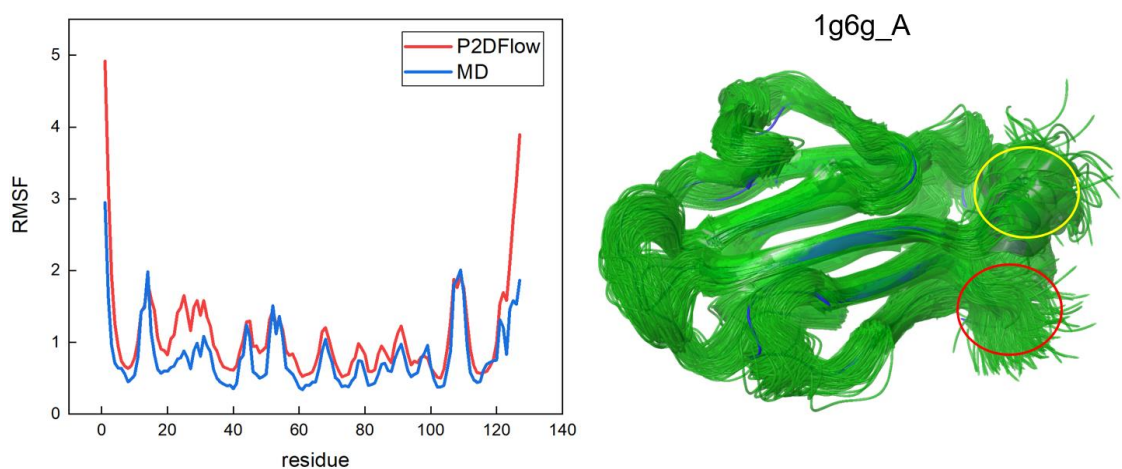

**Figure 7. RMSF of predicted conformations and that of MD conformations for 1g6g_A.** The experimental structures (blue) and all the sampled structures (green) are shown on the right. The sequence start (red circle) and end (yellow circle) are marked.

# 4. Conclusion

In this study, we introduce P2DFlow, a novel flow matching model designed to enhance the prediction of protein conformational ensembles. Our method integrates a specialized prior composed of ESMFold predictions with perturbations, introducing valuable biases, and includes an approximate energy mechanism to differentiate structures within conformational ensembles. P2DFlow not only predicts protein ensembles more swiftly than molecular dynamics simulations but also achieves higher accuracy than current methods like AlphaFlow and STR2STR, as confirmed by our thorough evaluation. By incorporating additional test cases with significant biological relevance, we have demonstrated the versatility and accuracy of our approach in capturing functionally distinct protein conformations.

To further refine model predictions and expand the diversity of the predicted conformational space, future work could explore two optimization avenues. Firstly, adopting a more precise prior, such as AlphaFold3 prediction, could provide a superior starting point for the sampling iteration process. Secondly, incorporating additional physical insights, like advanced force fields, would enhance the model's capacity to learn the correct distribution.

## Acknowledgements

This research work was supported by National Key Research and Development Program of China (2022YFC3400504), Shanghai Rising-Star Program (23QD1400600, 23YF1449200), Postdoctoral Fellowship Program of CPSF (GZC20232846, 2022M712402), National Natural Science Foundation of China (NSFC, No.22309134).

## References

1. Vögele, M., Zhang, B. W., Kaindl, J.; Wang, L., Is the functional response of a receptor determined by the thermodynamics of ligand binding? *Journal of Chemical Theory and Computation* 2023, 19 (22), 8414-8422.

2. Meller, A., Ward, M. D., Borowsky, J. H., et al., Predicting the locations of cryptic pockets from single protein structures using the PocketMiner graph neural network. *Biophysical journal* 2023, 122 (3), 445a.

3. Wu, R., Ding, F., Wang, R., et al., High-resolution de novo structure prediction from primary sequence. *BioRxiv* 2022, 2022.07. 21.500999.

4. Lu, J., Zhong, B., Zhang, Z.; Tang, J., Str2str: A score-based framework for zero-shot protein conformation sampling. *The Twelfth International Conference on Learning Representations* 2024.

5. Jumper, J., Evans, R., Pritzel, A., et al., Highly accurate protein structure prediction with AlphaFold. *Nature* 2021, 596 (7873), 583-589.

6. Lin, Z., Akin, H., Rao, R., et al., Evolutionary-scale prediction of atomic-level protein structure with a language model. *Science* 2023, 379 (6637), 1123-1130.

7. Lane, T. J., Protein structure prediction has reached the single-structure frontier. *Nature Methods* 2023, 20 (2), 170-173.

8. Ourmazd, A., Moffat, K.; Lattman, E. E., Structural biology is solved—now what? *Nature methods* 2022, 19 (1), 24-26.

9. Saldaño, T., Escobedo, N., Marchetti, J., et al., Impact of protein conformational diversity on AlphaFold predictions. *Bioinformatics* 2022, 38 (10), 2742-2748.

10. Chakravarty, D.; Porter, L. L., AlphaFold2 fails to predict protein fold switching. *Protein Science* 2022, 31 (6), e4353.

11. Shaw, D. E., Maragakis, P., Lindorff-Larsen, K., et al., Atomic-level characterization of the structural dynamics of proteins. *Science* 2010, 330 (6002), 341-346.

12. Torrie, G. M.; Valleau, J. P., Nonphysical sampling distributions in Monte Carlo free-energy estimation: Umbrella sampling. *Journal of Computational Physics* 1977, 23 (2), 187-199.

13. Laio, A.; Parrinello, M., Escaping free-energy minima. *Proceedings of the national academy of sciences* 2002, 99 (20), 12562-12566.

14. Noé, F., Olsson, S., Köhler, J.; Wu, H., Boltzmann generators: Sampling equilibrium states of many-body systems with deep learning. *Science* 2019, 365 (6457), eaaw1147.

15. Wayment-Steele, H. K., Ojoawo, A., Otten, R., et al., Predicting multiple conformations via sequence clustering and AlphaFold2. *Nature* 2024, 625 (7996), 832-839.

16. Stein, R. A.; Mchaourab, H. S., SPEACH_AF: Sampling protein ensembles and conformational heterogeneity with Alphafold2. *PLOS Computational Biology* 2022, 18 (8), e1010483.

17. Del Alamo, D., Sala, D., Mchaourab, H. S.; Meiler, J., Sampling alternative conformational states of transporters and receptors with AlphaFold2. *Elife* 2022, 11, e75751.

18. Zheng, S., He, J., Liu, C., et al., Towards predicting equilibrium distributions for molecular systems with deep learning. *arXiv preprint arXiv:2306.05445* 2023.

19. Lewis, S., Hempel, T., Jiménez Luna, J., et al., Scalable emulation of protein equilibrium ensembles with generative deep learning. *bioRxiv* 2024, 2024.12.05.626885.

20. Tang, Y., Yu, M., Bai, G., et al., Deep learning of protein energy landscape and conformational dynamics from experimental structures in PDB. *bioRxiv* 2024, 2024.06.27.600251.

21. Jing, B., Berger, B.; Jaakkola, T., AlphaFold Meets Flow Matching for Generating Protein Ensembles. *arXiv preprint arXiv:2402.04845* 2024.

22. Vander Meersche, Y., Cretin, G., Gheeraert, A., Gelly, J.-C.; Galochkina, T., ATLAS: protein flexibility description from atomistic molecular dynamics simulations. *Nucleic Acids Research* 2024, 52 (D1), D384-D392.

23. Lipman, Y., Chen, R. T., Ben-Hamu, H., Nickel, M.; Le, M., Flow matching for generative modeling. *International Conference on Learning Representations* 2023.

24. Chen, R. T., Rubanova, Y., Bettencourt, J.; Duvenaud, D. K., Neural ordinary differential equations. *Advances in neural information processing systems* 2018, 31.

25. Vignac, C., Krawczuk, I., Siraudin, A., et al., Digress: Discrete denoising diffusion for graph generation. *arXiv preprint arXiv:2209.14734* 2022.

26. Jing, B., Erives, E., Pao-Huang, P., et al., Eigenfold: Generative protein structure prediction with diffusion models. *arXiv preprint arXiv:2304.02198* 2023.

27. Stark, H., Jing, B., Barzilay, R.; Jaakkola, T. Harmonic prior self-conditioned flow matching for multi-ligand docking and binding site design. In *NeurIPS 2023 AI for Science Workshop*, 2023.

28. Albergo, M. S.; Vanden-Eijnden, E. J. a. p. a., Building normalizing flows with stochastic interpolants. *In The Eleventh International Conference on Learning Representations* 2022.

29. Chen, R. T.; Lipman, Y., Riemannian flow matching on general geometries. *arXiv preprint arXiv:2302.03660* 2023.

30. Yim, J., Campbell, A., Foong, A. Y., et al., Fast protein backbone generation with SE (3) flow matching. *arXiv preprint arXiv:2310.05297* 2023.

31. Pooladian, A.-A., Ben-Hamu, H., Domingo-Enrich, C., et al., Multisample flow matching: Straightening flows with minibatch couplings. *arXiv preprint arXiv:2304.14772* 2023.

32. Yim, J., Trippe, B. L., De Bortoli, V., et al., Se (3) diffusion model with application to protein backbone generation. *arXiv preprint arXiv:2302.02277* 2023.

33. Bose, A. J., Akhound-Sadegh, T., Huguet, G., et al., Se (3)-stochastic flow matching for protein backbone generation. *arXiv preprint arXiv:2310.02391* 2023.

34. Satorras, V. G., Hoogeboom, E.; Welling, M. E (n) equivariant graph neural networks. In *International conference on machine learning*, PMLR: 2021; pp 9323-9332.

35. Klein, L., Krämer, A.; Noé, F., Equivariant flow matching. *Advances in Neural Information Processing Systems* 2024, 36.

36. Shaul, N., Chen, R. T., Nickel, M., Le, M.; Lipman, Y. On kinetic optimal probability paths for generative models. In *International Conference on Machine Learning*, PMLR: 2023; pp 30883-30907.

# Table of Contents

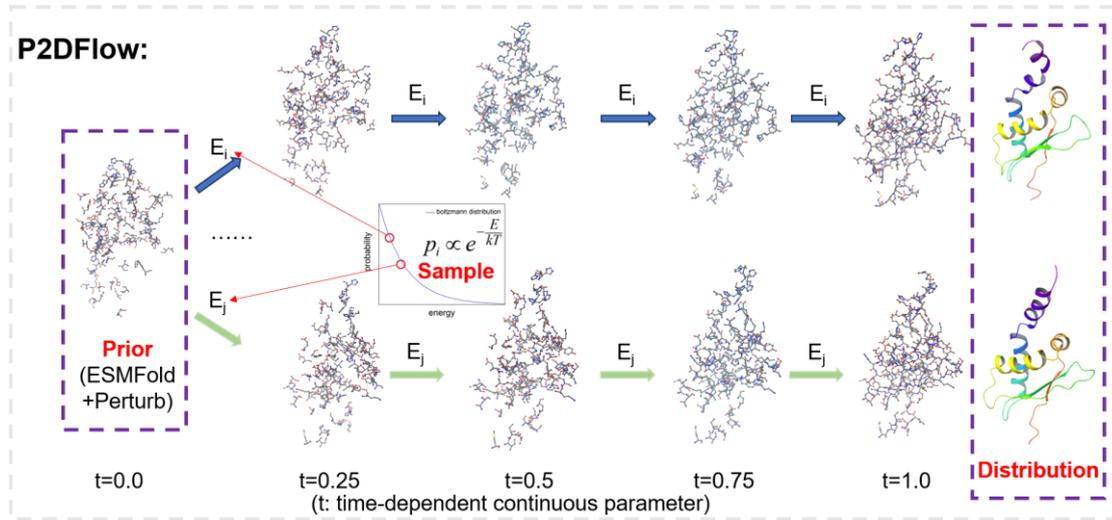